# Direct STM Measurements of R-type and H-type Twisted MoSe$_2$/WSe$_2$

Rachel Nieken[1], Anna Roche[1], Fateme Mahdikhanysarvejahany[1], Takashi Taniguchi[2], Kenji Watanabe[3], Michael R. Koehler[4], David G. Mandrus[5-7], John Schaibley[1], Brian J. LeRoy[1]

[1]Department of Physics, University of Arizona, Tucson, Arizona 85721, USA
[2]International Center for Materials Nanoarchitectonics, National Institute for Materials Science, 1-1 Namiki, Tsukuba 305-0044, Japan
[3]Research Center for Functional Materials, National Institute for Materials Science, 1-1 Namiki, Tsukuba 305-0044, Japan
[4]JIAM Diffraction Facility, Joint Institute for Advanced Materials, University of Tennessee, Knoxville, TN 37920
[5]Department of Materials Science and Engineering, University of Tennessee, Knoxville, Tennessee 37996, USA
[6]Materials Science and Technology Division, Oak Ridge National Laboratory, Oak Ridge, Tennessee 37831, USA
[7]Department of Physics and Astronomy, University of Tennessee, Knoxville, Tennessee 37996, USA

**When semiconducting transition metal dichalcogenides heterostructures are stacked the twist angle and lattice mismatch leads to a periodic moiré potential. As the angle between the layers changes, so do the electronic properties. As the angle approaches 0- or 60-degrees, interesting characteristics and properties such as modulations in the band edges, flat bands, and confinement are predicted to occur. Here we report scanning tunneling microscopy and spectroscopy measurements on the band gaps and band modulations in MoSe$_2$/WSe$_2$ heterostructures with near 0 degree rotation (R-type) and near 60 degree rotation (H-type). We find a modulation of the band gap for both stacking configurations with a larger modulation for R-type than for H-type as predicted by theory. Furthermore, local density of states images show that electrons are localized differently at the valence band and conduction band edges.**

Stacking two monolayer van der Waals materials with either a lattice mismatch or a twist angle between them creates a moiré superlattice which changes the electronic structure of the heterostructure[1-8]. For example, in twisted bilayer graphene the twist angle controls the emergence of charged ordered states[9], unconventional superconductivity[7], correlated insulating states[8] and magnetic states[10], which are the product of the formation of moiré potentials in the structure. In a case of a lattice mismatch between the layers such as graphene and boron nitride heterostructures, new Dirac cones and the Hofstadter butterfly pattern appear even at zero twist angle[2,4-7]. Novel quantum phenomenon have also been predicted and observed in twisted transition metal dichalcogenide (TMD) heterostructures including homobilayers[11] and heterobilayers[12-18]. Near R- or H-type stacking, both the electronic and topographic structure are modified leading to flat bands, confinement, and atomic reconstruction[1,11-14,19-23]. TMD-TMD bilayers are of interest due to the larger range of angles at which flat bands can form and TMD-TMD heterostructures can more accurately represent the Hubbard model than graphene[24-26].

One way to measure the electronic properties of these structures is through scanning tunneling microscopy (STM) and spectroscopy (STS), which unlike far-field optical measurements can provide direct measurements of the magnitude and wavelength of the moiré potential as well as the electronic structure. STM uses a sharp metal tip to directly probe the surface by applying a bias voltage between the tip and the sample. Since STM probes the electronic properties of the surface, this requires that the surface of the devices be exposed and clean which is challenging for TMDs. However, recently a few STM, conductive atomic force microscopy (cAFM), transmission electron microscopy (TEM), and scanning TEM (STEM) observations have been conducted. STM measurements on twisted bilayer $WSe_2$ found flat bands near both 0 and 60 degrees and that the flat bands are localized differently for the two cases[20]. Other experiments on heterobilayer TMDs include observations of a modulation in the moiré potential at a moiré length of 8.3 nm in chemical vapor deposition grown $MoS_2/WSe_2$ heterostructures[19] and variations in the spectral peaks' sharpness between 80 K and 5 K[1]. Experimental results for STM measurements on $WSe_2/WS_2$ heterobilayers reveal flat bands near the edge of the valence edge band at low temperature[21].

Lastly, for $MoSe_2/WSe_2$ heterostructures recent STM, cAFM, and TEM measurements at room temperature mapped the variation in the energy of the valence and conduction band of the H-type (60 degree) $MoSe_2/WSe_2$ structure for wavelengths ranging from 6 nm to 17 nm and obtained tunneling spectroscopy at the high symmetry stacking locations. These measurements found that there was a decrease in the magnitude of the modulation as the moiré wavelength decreased[27]. STEM and TEM measurements on $MoSe_2/WSe_2$ explored atomic reconstruction of both types at room temperature for CVD grown structures for twist angles less than 1 degree[22,23].

In this work, we measure both R-type (near 0 degree) and H-type (near 60 degree) $MoSe_2/WSe_2$ heterostructures to understand modulations in the band structure with respect to the moiré pattern. Using STS measurements, we map the band edges and modulation of the band gap for a range of twist angles at 77 K and 4.5 K. A schematic of a typical STM device and the experimental setup is shown in Fig. 1a. Optical microscope images of the R-type and H-type devices are shown in Supplementary Figs. 2a and b respectively. For the experiments conducted in this work the $WSe_2$ is the top layer and thus the tungsten STM tip preferentially tunnels into this layer. One of the challenges of performing the STM measurements of TMDs at low-temperature is making good electrical contact to the device. We found that placing a graphene layer under the TMDs allowed for reliable STM measurements over a broad range of tunnel voltages. Without the graphene layer when probing energies near or inside the band gap, the tip could crash into the sample destroying the sample and the tip in the process. To address the challenges of obtaining clean TMD heterostructures, devices were annealed in vacuum and AFM cleaned before beginning STM measurements (see the Methods section for more details). The heterostructures were then scanned in the STM to find clean, defect-free areas before taking spectroscopy measurements.

When TMD heterostructures such as MoSe$_2$/WSe$_2$ are twisted, the atomic stackings between the layers vary periodically in space. There are three main high symmetry points that are of interest as shown in Figs. 1d and e along with the corresponding atomic registry stackings and moiré lattices. For the R-type stacking the three high symmetry atomic stacking orders are tungsten on molybdenum ($R_{M'}^{M}$) or equivalently selenium on selenium ($R_{X'}^{X}$), tungsten on selenium ($R_{X'}^{M}$), and selenium on molybdenum ($R_{M'}^{X}$), where M is the metal atom, X is the chalcogen atom, the apostrophe denotes the bottom layer, and the R indicates the R-type configuration. The H-type also has three high symmetry stacking configurations $H_{X'}^{M}$ ($H_{M'}^{X}$), $H_{M'}^{M}$, and $H_{X'}^{X}$ labeled in Fig. 1e. In Fig. 1b, we show STM topography for a 5 nm moiré in R-type heterostructure at 4.5 K and Fig. 1c shows a 12 nm moiré in an H-Type device. The R-type heterostructure has a hexagonal moiré pattern while the H-type has a triangular pattern in agreement with previous results and calculations[22,23,27].

To compare the electronic properties of the R- and H- type stacking configurations, we perform scanning tunneling spectroscopy measurements at the high symmetry locations of the moiré unit cells. In Fig. 2, we compare constant height point spectroscopy of the R-type and H-type structures at 77 K. The $R_{X'}^{M}$ (green) stacking has the smallest band gap at 2.03 +/-0.01 eV while the largest band gap of 2.29 +/-0.01 eV occurs at the $R_{M'}^{M}$ (red) site, and the $R_{M'}^{X}$ (black) site has a band gap of 2.08 +/-0.02 eV. Therefore, the maximum modulation of the band gap is around 260 meV for this stacking configuration and moiré wavelength. For the R-type heterostructures, we also observed that most of the modulation in the band gap occurs from a shift in the valence band. For the H-type the smallest band gap occurs at the $H_{M'}^{M}$, (blue) stacking site which has a band gap of 1.64 +/-0.01 eV. The largest band gap corresponding to the $H_{X'}^{X}$ (red) site is 1.80 +/-0.01 eV. The $H_{X'}^{M}$ has a band gap of 1.76 +/-0.01 eV. Thus, the H-type device shows modulation of both band edges with a modulation of 160 meV for this moiré wavelength. Comparing the modulation in the R-type and the H-type we note that the R-type modulation is 1.6 times larger than the H-type which is smaller than the theoretically predicted 3:1 ratio although the difference may be caused by the different moiré wavelengths[28].

From the line spectroscopy of the R-type device taken at 77 K in Fig. 3b, we observe a consistent oscillation in the band energy at the valence band edge. The STM topography in Fig. 3a shows that the oscillation corresponds to the 5 nm moiré period between the high symmetry sites. The bright spot in the topography ($R_{M'}^{M}$) aligns with the local minima in the valence band with the band edge at around -1.51 +/- 0.01 eV. The other two sites align with the two peaks in the valence band with a local maxima value of -1.31 +/- 0.01 eV for the $R_{M'}^{X}$ (black) and a global maxima value of -1.25 +/- 0.01 eV for the $R_{X'}^{M}$ (green) site. Spectroscopy of the conduction band edge showed little to no variation with at most a slight dip at the $R_{M'}^{M}$ site (Supplementary Fig. 5), which can also be seen in Fig 2a. In Fig. 1f, we see the alignment between the WSe$_2$ and MoSe$_2$ is a type II band alignment such that the MoSe$_2$ is the conduction band edge and the WSe$_2$ is the valence band edge. Therefore, for R-type devices the oscillation in the band gap mainly arises from the valence band edge of the WSe$_2$. By taking the difference in the global minima and maxima in the valence band we find the maximum amplitude in oscillation is 260 meV and the difference between the two maxima to be 60 meV. Using the constant value from the conduction band edge, we find that the band gaps for each of the three high symmetry points are 2.28 +/- 0.01 eV, 2.07 +/- 0.01 eV, and 2.01 +/- 0.01 eV respectively. We note that the band gaps are comparable with the spectroscopy taken at another site on the sample. In both the point spectroscopy and the line spectroscopy the maximum band modulation occurs at a different stacking configuration compared to the DFT calculations which predict a maximum at $R_{X'}^{M}$[28]. However, these calculations do not consider lattice relaxations effects. It has been shown for H-type stacking the inclusion of relaxation effects in the DFT model changes which stacking order has the largest band gap[27]. Further, reports have shown that strain and interlayer distance can affect the band modulation[29].

In Figs. 3d-f, LDOS images for a R-type device at 4.5 K are shown. At energies near the valence band edge around -1.3 eV the electronic wave functions are localized between the black and green sites (Fig. 3d). As the energy decreases to -1.7 eV the electrons become localized on the black site (Fig. 3e) and as the energy decreases farther to -1.8 eV the electrons localize on the red site (Fig. 3f). This is in good agreement with the line spectroscopy in Fig. 3b. From the color scale in the line spectroscopy, we note that the dI/dV values are higher at the black and green sites at higher energy and as the energy decreases the dI/dV becomes highest on the red site as verified by the LDOS images.

Next, we focus more closely on the H-type device. In Fig. 4b, the constant current line spectroscopy at 77 K for the H-type device is shown. The valence band varies with stacking configuration, but unlike the R-type device the conduction band edge also varies with position. By taking the difference between the conduction and a valence band edge at each of the lines in Fig. 4b we find the band gap and determine the modulation between the sites. The high symmetry point corresponding to $H_{M'}^{M}$ (blue) stacking has a band gap of 1.65 +/- 0.01 eV, the $H_{X'}^{X}$ (red) stacking has a band gap of 1.81 +/- 0.01 eV, and the band gap at the $H_{X'}^{X}$ (green) stacking is 1.78 +/- 0.01 eV. The modulation between the $H_{M'}^{M}$ site and the $H_{X'}^{X}$ site is 160 meV and the difference between the $H_{X'}^{X}$ and the $H_{X'}^{M}$ site is determined to be 30 meV. We note that the band gaps at each of the sites and the modulations in the band are comparable to the gaps and modulation determined from the point spectroscopy in Fig. 2b and in the two-dimensional grid spectroscopy in Supplementary Fig. 3. Further, the small modulation between the $H_{X'}^{X}$ and $H_{X'}^{M}$ aligns with previous work where the two bands were observed to decrease in separation as the moiré wavelength decreased from 13 nm to 6 nm [27].

LDOS images for an H-type device were taken on a 12 nm area at 4.5 K. In Figs. 4d-g the LDOS progresses from the lowest valence band to highest conduction band energy. At low energy, electrons are localized on the $H_{X'}^{M}$ and $H_{X'}^{X}$ sites in the unit cell. As the energy nears the valence band edge the electrons move to localize on the $H_{M'}^{M}$ site. This is in good agreement with the line spectroscopy in Fig. 4b. The dI/dV spectroscopy is largest near the valence band edge around the $H_{M'}^{M}$ site and as the energy decreases to around -1.3 eV the dI/dV spectroscopy is smallest at the $H_{M'}^{M}$ site. The difference in energy between the line spectroscopy and LDOS images is due a slight change in the tip conditions and all measurements are offset to a lower energy by about 0.2 eV. For the conduction band edge shown in Fig. 4f, the electrons appear to be localized around the edges of the moiré unit cell. As the energy increases in the conduction band the electron density is highest around the $H_{M'}^{M}$' and $H_{X'}^{X}$ sites and exhibit a triangular wave function. Further intermediate LDOS maps between 0.675 eV and 0.8 eV for the conduction band and intermediate LDOS maps between -1.275 eV and -1.6 eV for the valence band which present the progression of the wave functions are shown in supplementary Fig. 4.

In conclusion, our spectroscopic measurements of the R-type stacking show that the modulation in the band between the maximum and minima is larger than previous predictions and the largest modulation in the bands occurs at the $R_{M'}^{M}$ site. The modulation in the bands is dominated by shifts in the valence band for the R-type while both bands are strongly modulated for H-type stacking. The R-type also has overall larger band gaps than the H-type. Our H-type results agree with previous measurements taken at room temperature with signs of localization near the conduction and valence band edges of the H-type device.

**Methods**

*Device Fabrication.* Monolayer MoSe$_2$ and WSe$_2$, 10 to 30 nm thick hexagonal boron nitride, and monolayer or bilayer graphene were obtained through mechanical exfoliation techniques from bulk crystals[30]. The alignment between the MoSe$_2$ and WSe$_2$ was determined through second harmonic generation (SHG)[31]. A polycarbonate film was used to pick up the pieces in order of WSe$_2$, MoSe$_2$, graphene, and hBN. The film was melted onto the silicon chip and removed with chloroform, Isopropanol (IPA), and dried with nitrogen gas.

*Contacts.* A bilayer of 495 and 950 PMMA was spin coated on the sample at 2000 rpm and cured using a hot plate. Electron beam lithography was used to define the contact pattern followed by thermal evaporation of 10 nm of chrome and 50 nm of gold. The PMMA was dissolved with acetone and IPA before a final annealing of the sample.

*Annealing.* The heterostructure was annealed in vacuum at 250 degrees Celsius after the mechanical transfer and SHG measurements. A 250-degree vacuum anneal was used because the process removes contaminants obtained from the transfer process while being less aggressive and rough on the device than higher temperatures. The high temperature of the anneal has the possibility to align the layers into either the R- or H-type stacking configuration and therefore give a long moiré wavelength. Adding metal contacts to the device tends to hold the layers in place. Therefore, gold contacts were added after annealing allowing the possibility of the TMD layers to align themselves during annealing.

*AFM Cleaning*. AFM cleaning was conducted after the first anneal. An AFM tip was used in contact mode at a force of 80 nN to push away residues on the surface of the heterostructure. The force used was determined by starting with the smallest force and increasing the force in 1 nN steps until the scanned area was devoid of larger bubbles and other dirt. The minimal amount of force was used to avoid tearing, harming, and changing the device. In the AFM cleaned samples, the STM imaged relatively uniform moiré wavelengths across the sample indicating that the moiré was still intact and the cleaning process did not introduce strain. There is a possibility that the cleaning process changes the wavelength since it may move the materials with respect to each other. However, the SHG measurements shown in Supplementary Figure 2, which are taken before AFM cleaning, show the same twist angle as measured by STM. This suggests that the AFM cleaning procedure does not affect the alignment of the two TMDs.

*STM Measurements*. STM and STS measurements were obtained at either 77 K or 4.5 K in ultra-high vacuum conditions. A tungsten tip was used to probe the heterostructure with the $WSe_2$ layer on the top. A lock-in amplifier was used to perform the dI/dV spectroscopy measurements.

*SHG Measurements*. Second harmonic generation (SHG) measurements were used to determine the crystal orientations of the $WSe_2$ and $MoSe_2$ monolayers. Linearly polarized laser (120 fs model-locked Ti:sapphire) light at 800 nm was focused onto the monolayer with a 50x objective and passed through a half wave plate. The SHG signal at 400 nm was collected by a spectrometer. The monolayers were aligned by extracting the angle where the SHG signal of the $MoSe_2$ and $WSe_2$ monolayers are at a maximum. After the mechanical transfer, SHG was used again on the isolated $MoSe_2$, $WSe_2$, and overlap regions to determine if the resulting angle was near 0 or near 60 degrees. SHG data for the R and H type devices are shown in Supplementary Figs. 2e and f.

**Data Availability**

Data that support the plots within this paper and other findings of this study are available from the corresponding author upon reasonable request.

**Supplementary Material**

See supplementary material for a description of the averaging procedure used to extract the band gap from spectroscopy measurements. There are also figures showing optical microscopy images of the devices along with high-resolution STM topography and second harmonic generation measurements for the alignment of the TMDs. In addition, there are two-dimensional grid spectroscopy measurements and LDOS images for an H-type device and line spectroscopy measurements for a R-type device.

**Acknowledgments**

This work is supported at the University of Arizona by the Army Research Office under Grant nos. W911NF-18-1-0420 and W911NF-20-1-0215 and the National Science Foundation under grant Nos. DMR-1708406, DMR-1838378, and DMR-2054572. DGM acknowledges support from the Gordon and Betty Moore Foundation's EPiQS Initiative, Grant GBMF9069. K.W. and T.T. acknowledge support from the Elemental Strategy Initiative conducted by the MEXT, Japan

(Grant Number JPMXP0112101001) and JSPS KAKENHI (Grant Numbers 19H05790, 20H00354 and 21H05233).

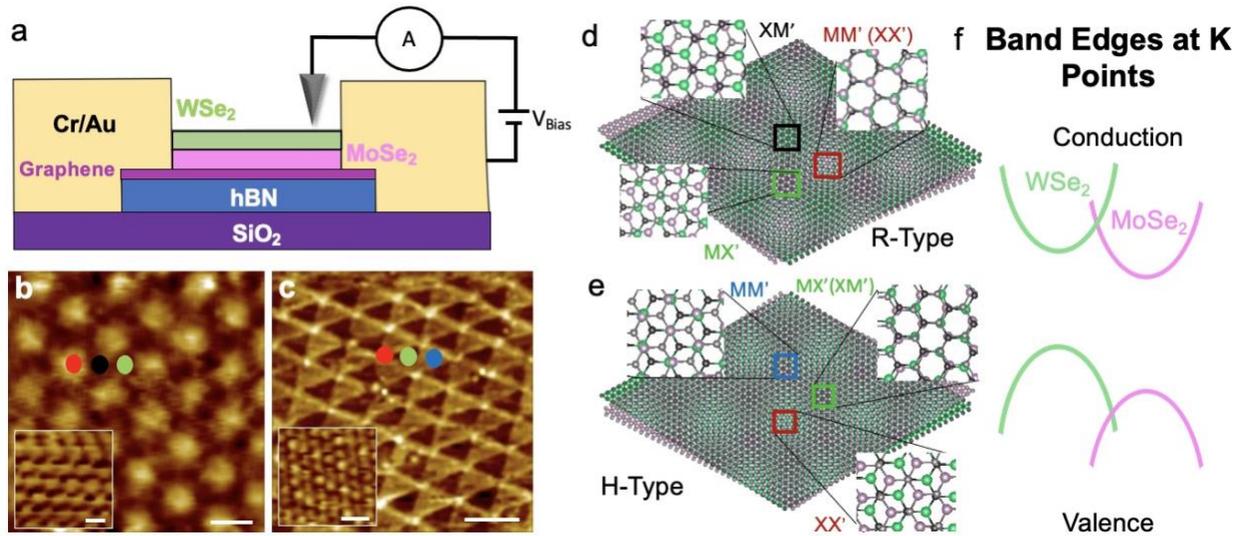

**Figure 1**. *Experimental Setup and Structure of Twisted MoSe$_2$/WSe$_2$*. a) Schematic of MoSe$_2$/WSe$_2$ heterostructure with the following layers in order from top to bottom: monolayer WSe$_2$, monolayer MoSe$_2$, monolayer or bilayer graphene as a conducting layer, and 10-30 nm hBN. Scanning tunneling microscopy and spectroscopy is performed using an atomically sharp tungsten tip with a bias voltage, $V_{bias}$, between the tip and the heterostructure. b) STM topography image of an R-type heterostructure with a 5 nm moiré taken at $V_{bias}$ = -3.0 V, $I_{set}$ = 100 pA, and 4.5 K. The inset is an STM atomic resolution image taken at $V_{bias}$ = -2.8 V, $I_{set}$ = 200 pA. The white scale bar is 5 nm for the main image and 3 Å for the inset. c) STM topography image of an H-type heterostructure with a 12 nm moiré taken at $V_{bias}$ = -2.7 V, $I_{set}$ = 200 pA, and 4.5 K. The inset is a STM atomic resolution image taken at $V_{bias}$ = -2.7 V, $I_{set}$ = 200 pA. The white scale bar is 10 nm for the main image and 6 Å for the inset. d) A schematic of the moiré for a R-Type (near zero degree) twisted MoSe$_2$/WSe$_2$ structure. e) A schematic of the moiré for an H-Type (near 60-degree) twisted MoSe$_2$/WSe$_2$ structure. d, e) The atomic stacking configurations for each of the three high symmetry sites is shown and labeled, where M is Tungsten (green), M' is Molybdenum (pink), X, X' are Selenium in the top (grey) and bottom (black) layer. f) An illustration of the type II band alignment between the WSe$_2$ (green) and MoSe$_2$ (pink) bands at the K-point.

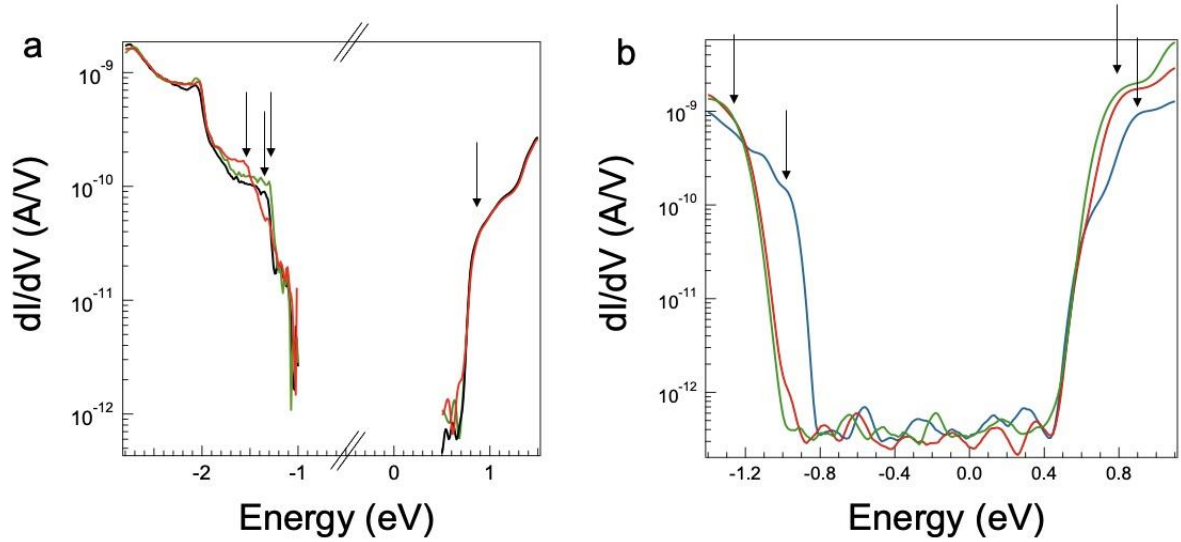

**Figure 2.** *Point Spectroscopy.* a) STS constant height point spectroscopy of the R-Type device on the $R_{M'}^{M}$ ($R_{X'}^{X}$) (red), $R_{M'}^{X}$ (black), and $R_{X'}^{M}$ (green) symmetry sites for a 5 nm moiré at 4.5 K with at set voltage $V_{set}$ = -2.8 V for the valence band ($V_{set}$ = -1.5 V for the conduction band) and current $I_{set}$ = 200 pA. b) STS constant height point spectroscopy for the H-Type device on the $H_{M'}^{M}$(blue), $H_{X'}^{M}$ (green), and $H_{X'}^{X}$ (red) symmetry sites for a 5 nm moiré at 77 K with a set voltage $V_{set}$ = -1.4 V and current $I_{set}$ = 500 pA. a), b) The black arrows indicate the locations where the band edge values were extracted.

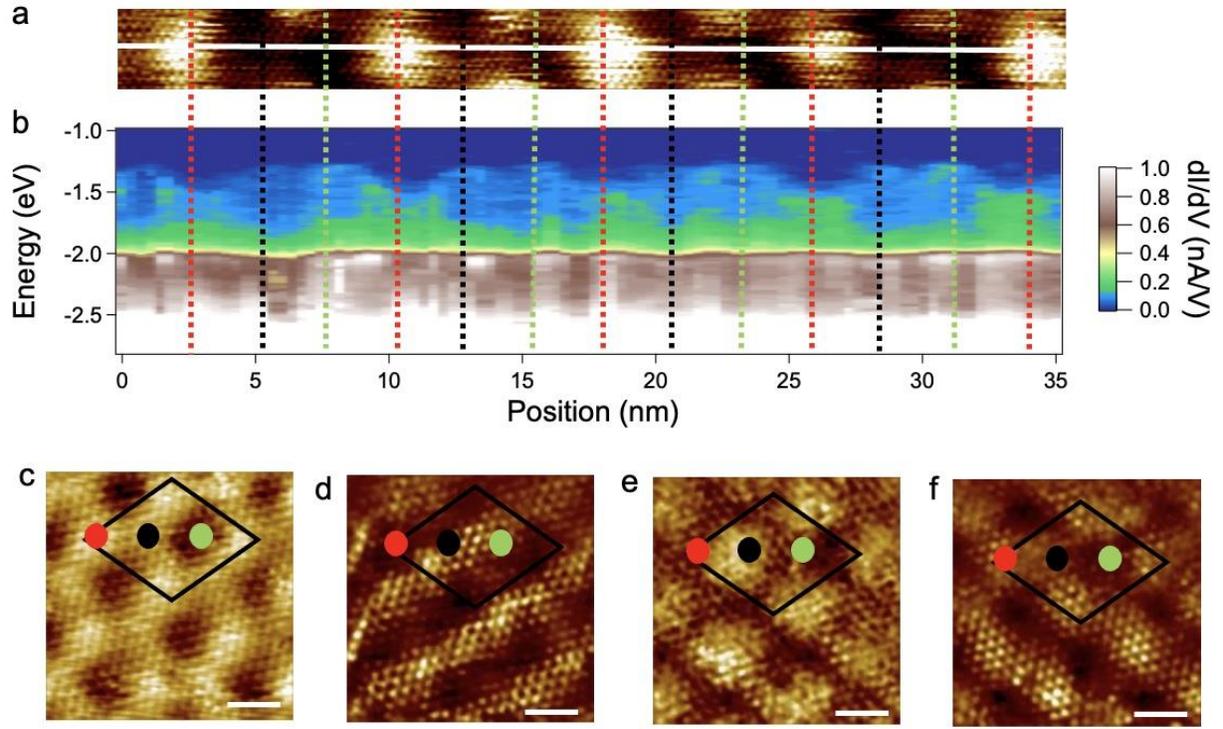

**Figure 3.** *R-Type Valence Band Oscillations and LDOS.* a) STM topography showing the moiré pattern. The white line indicates the line cut for the spectroscopy. b) STS constant height line spectroscopy of the R-Type moiré at 77 K is shown. The $R^M_{M'}$ ($R^X_{X'}$), $R^X_{M'}$, and $R^M_{X'}$ are denoted by the red, black, and green lines, respectively. c) A 15 nm by 15 nm STM topography image taken at $V_{bias}$ = -1.8 V and $I_{set}$ = 200 pA. d), e) ,f) LDOS images corresponding to the topography in c) at $V_{bias}$ = -1.3 V, $V_{bias}$ = -1.7 V, and $V_{bias}$ = -1.8 V, respectively. For all LDOS images $I_{set}$ = 200 pA. The LDOS were taken at 4.5 K. The black diamond marks the unit cell. The red, black, and green dots indicate the $R^M_{M'}$ ($R^X_{X'}$), $R^X_{M'}$, and $R^M_{X'}$ sites, respectively. The scale bar in all images is 3 nm.

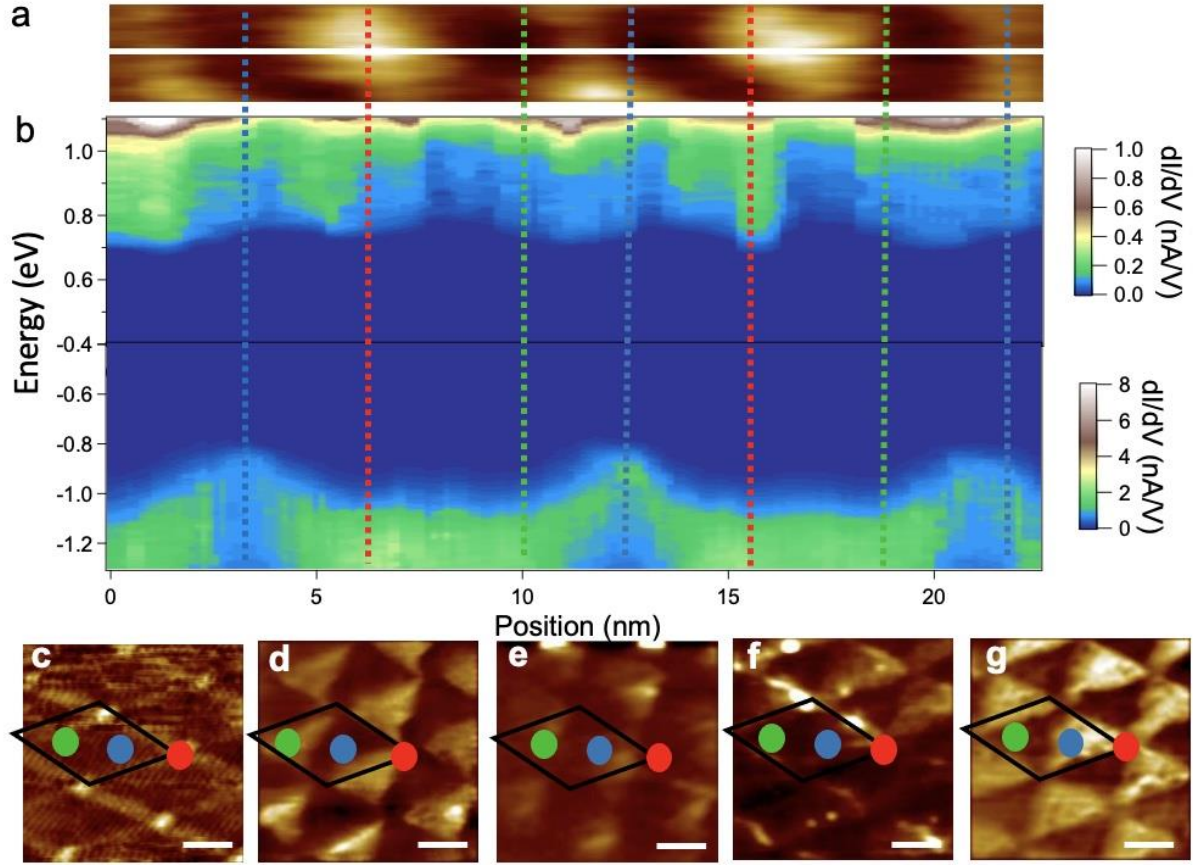

**Figure 4.** *H-Type Band Oscillations and LDOS.* a) STM topography image taken at $V_{bias}$ = -1.3 V and 77 K. The white line denotes the location of the spectroscopy in b. b) STS line spectroscopy of an H-Type device with 5 nm moiré at $V_{bias}$ = -1.3 V, $I_{set}$ = 200 pA, and a temperature of 77 K. The $H^M_{M'}$, $H^M_{X'}$ ($H^X_{M'}$), and $H^X_{X'}$ are denoted by the blue, green, and red lines, respectively. c) A 45 nm by 45 nm STM topography image taken at $V_{bias}$ = -2.8 V and $I_{set}$ = 200 pA of a 12 nm moiré. d), e), f), and g) LDOS images at $V_{bias}$ = -1.5 V, $V_{bias}$ = -1.275 V, $V_{bias}$ = 0.675 V, $V_{bias}$ = 0.8 V respectively corresponding to the area shown in (c). For all LDOS images $I_{set}$ = 200 pA. The black diamond marks the unit cell with the $H^X_{X'}$ at the four corners. The $H^M_{M'}$, $H^M_{X'}$ ($H^X_{M'}$), and $H^X_{X'}$ are denoted by the blue, green, and red dots, respectively. The images were taken at 4.5 K. The white scale bar in all images is 10 nm.

# Supplementary Materials and Figures

# Direct STM Measurements of R-type and H-type Twisted MoSe$_2$/WSe$_2$

Rachel Nieken[1], Anna Roche[1], Fateme Mahdikhanysarvejahany[1], Takashi Taniguchi[2], Kenji Watanabe[3], Michael R. Koehler[4], David G. Mandrus[5-7], John Schaibley[1], Brian J. LeRoy[1]

[1]Department of Physics, University of Arizona, Tucson, Arizona 85721, USA
[2]International Center for Materials Nanoarchitectonics, National Institute for Materials Science, 1-1 Namiki, Tsukuba 305-0044, Japan
[3]Research Center for Functional Materials, National Institute for Materials Science, 1-1 Namiki, Tsukuba 305-0044, Japan
[4]JIAM Diffraction Facility, Joint Institute for Advanced Materials, University of Tennessee, Knoxville, TN 37920
[5]Department of Materials Science and Engineering, University of Tennessee, Knoxville, Tennessee 37996, USA
[6]Materials Science and Technology Division, Oak Ridge National Laboratory, Oak Ridge, Tennessee 37831, USA
[7]Department of Physics and Astronomy, University of Tennessee, Knoxville, Tennessee 37996, USA

# Extraction of Band Gap Values

The point spectroscopies in Fig. 2 are an average of 20 repeated measurements on each of the different high symmetry sites for both the R- and H-type devices. An example of this data is shown in supplementary Fig. 1 for the $R_{M'}^{M}$ (red) site. The edge of the band was determined by extracting the points where the slope of the spectroscopy changes. This was done for both the conduction and valence band for each spectroscopy curve. Then, the values were averaged to determine the energy of the band edge in this case $-1.528 \pm 0.013$ eV for the valence band and $0.760 \pm 0.005$ eV for the conduction band. The band gap, $2.288 \pm 0.014$ eV was determined as the difference in energy between the conduction and valence band edges. The errors in the band edges and gaps were found by taking the standard deviation of the individual measured values. This gave a typical error of 10 meV for the band gap at the different locations. For the line spectroscopy in Figures 3 and 4, a similar method was used but the spectroscopy curves were averaged over a 1-nm area near each high symmetry point. The resulting spectroscopy at each equivalent repeating symmetry point was analyzed to determine the band edges and band gaps. The repeating symmetry points were then averaged to obtain the reported values and associated errors.

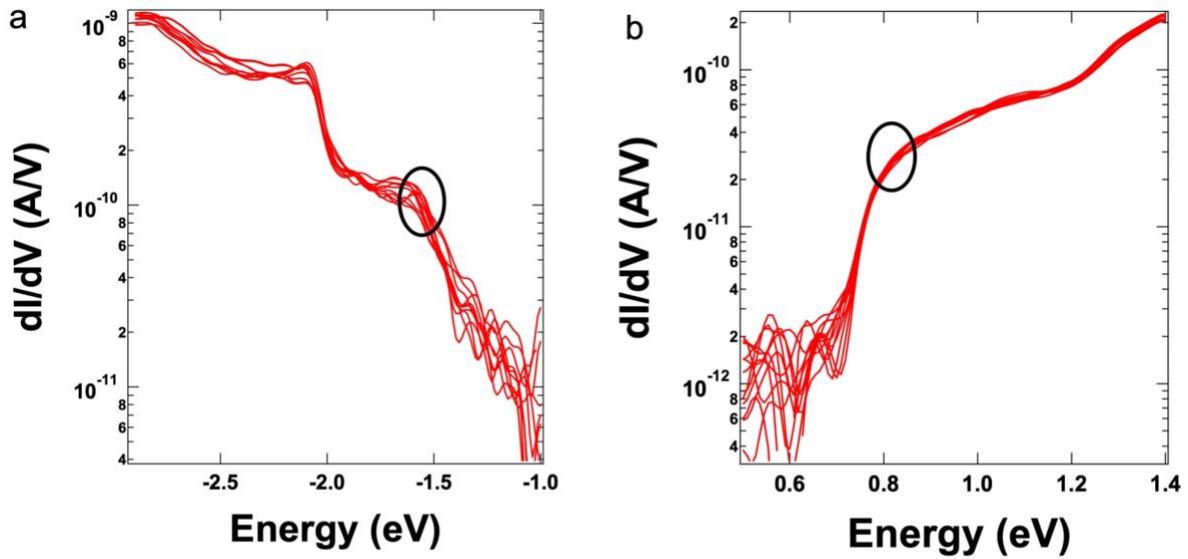

**Supplementary Figure 1.** *Band Gap Calculations.* a) Constant height point spectroscopy of the valence band of the R-Type device at the $R^M_{M'}$ ($R^X_{X'}$) (red) symmetry site for a 5 nm moiré at 77 K with at set voltage $V_{set}$ = -2.8 V and current $I_{set}$ = 200 pA. b) Constant height point spectroscopy of the conduction band at the $R^M_{M'}$ ($R^X_{X'}$) (red) symmetry site with at set voltage $V_{set}$ = -1.5 V and current $I_{set}$ = 200 pA. The circles indicate the approximate energy of the band edges.

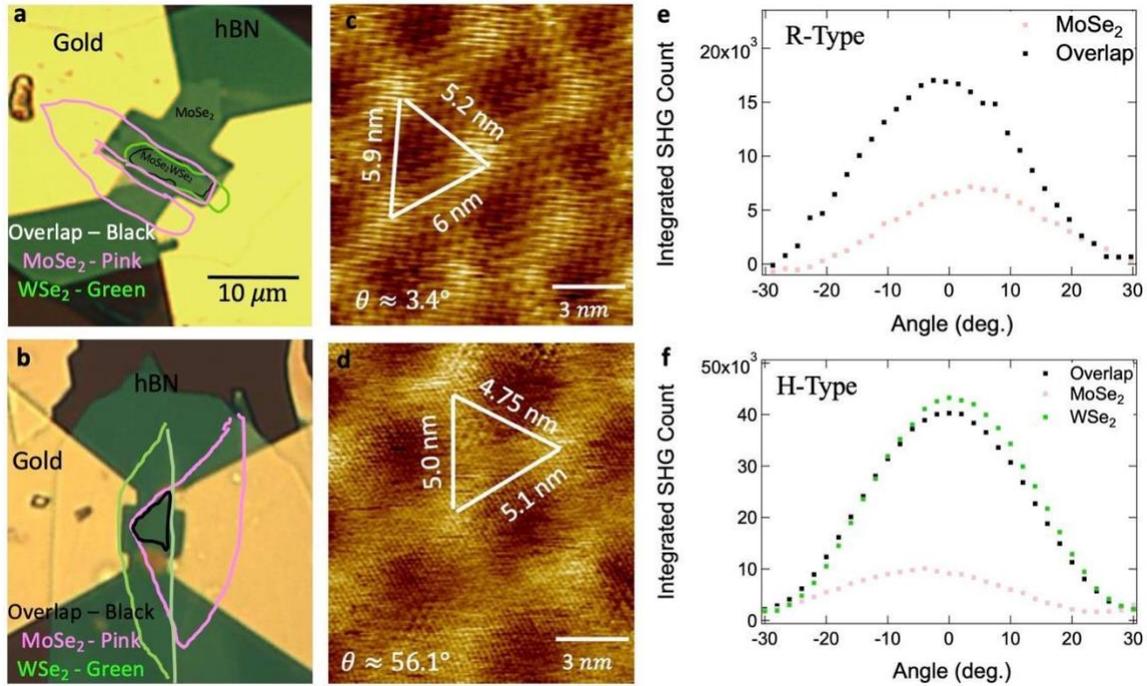

**Supplementary Figure 2.** *Devices, Angles, and SHG.* a) Optical microscope image of the R-Type device with each of the layers and the overlap region outlined. b) A moiré STM topography image taken at $V_{bias}$ = -2.1 V and $I_{set}$ = 200 pA of the R-type device with each of the different moiré lattice vectors measured and the corresponding twist angle. c) SHG of the R-type overlap, $MoSe_2$ regions. No isolated monolayer $WSe_2$ was present after transfer. d) Optical microscope image of the H-type device with each of the layers and the overlap region outlined. e) A moiré STM topography image taken at $V_{bias}$ = -2.3 V and $I_{set}$ = 200 pA of the H-type device with each of the different moiré lattice vectors measured and the corresponding twist angle. f) SHG of the H-type overlap, $MoSe_2$ region, and $WSe_2$ region.

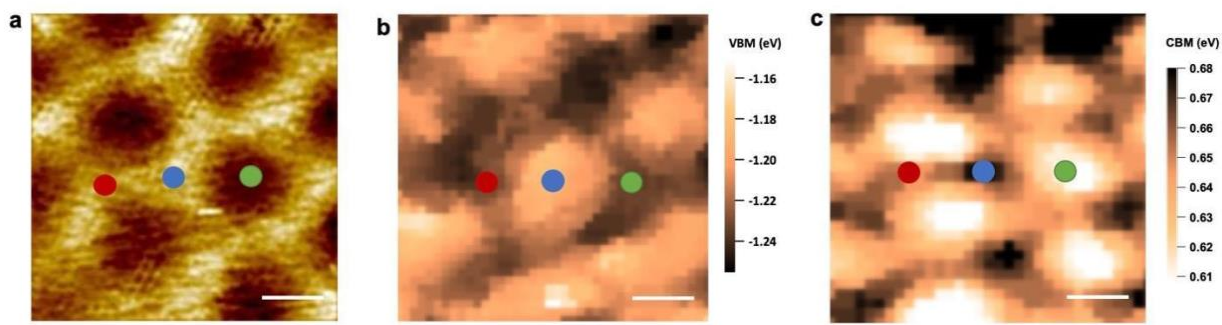

**Supplementary Figure 3.** *H-Type Energy Modulation.* a) STM topography taken at $V_{bias}$ = -2.9 V, $I_{set}$ = 200 pA, and 77 K for the 5 nm H-type device. b) Energy modulation in the valence band taken 20 nm away from a) and c), where the color scale is the band maxima in eV. c) Energy modulation in the conduction band, where the color scale is the band minima in eV. The white scale bar in all images is 3 nm.

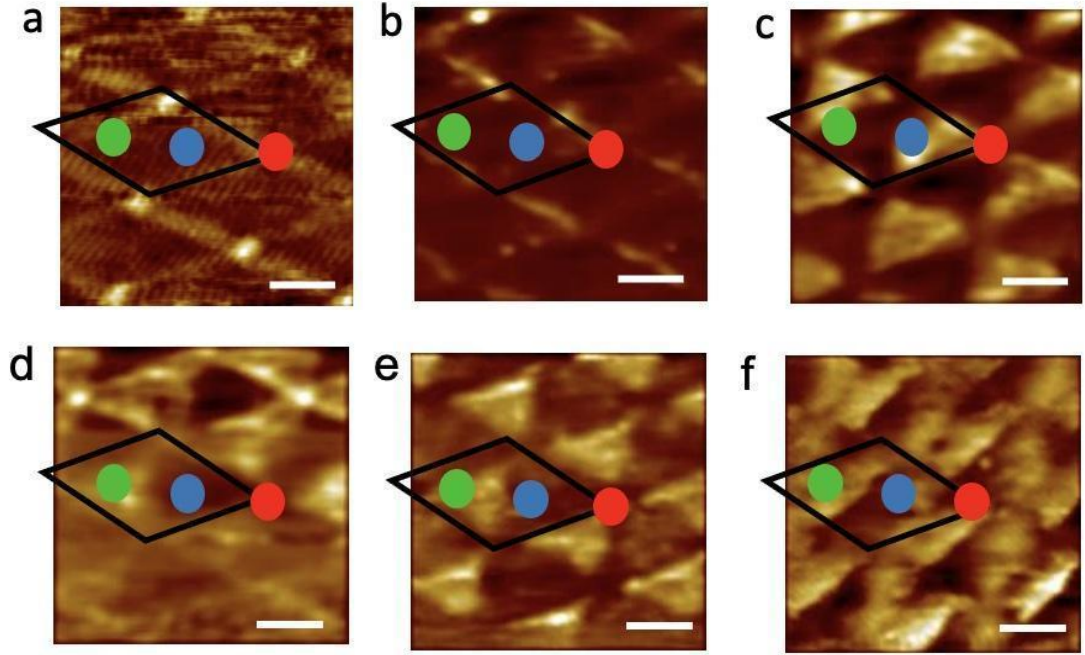

**Supplementary Figure 4.** *Additional LDOS images.* a) A 45 nm by 45 nm STM topography image taken at $V_{bias} = -2.8$ V and $I_{set} = 200$ pA of the 12 nm moiré at 4.5 K. b), c), d), e), f) STM LDOS images at $V_{bias} = 0.7$ V, $V_{bias} = 0.75$ V, $V_{bias} = -1.6$ V, $V_{bias} = -1.4$ V, and $V_{bias} = -1.3$ V respectively are shown. All LDOS correspond to the topography in a) and in all LDOS image $I_{set} = 200$ pA and the scans were taken at 4.5 K. The black diamond marks the unit cell with the $H_{X'}^{X}$ as the four corners. The $H_{M'}^{M}$, $H_{X'}^{M}$ ($H_{M'}^{X}$), and $H_{X'}^{X}$ are denoted by the blue, green, and red dots, respectively. The white scale bars denote 10 nm.

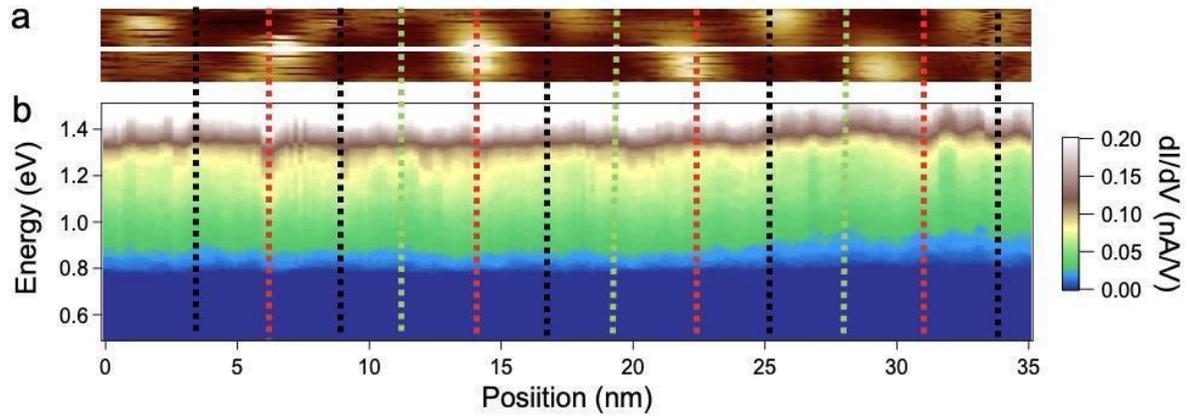

**Supplementary Figure 5.** *R-Type Conduction Band.* a) STM Topography at $V_{bias}$ = -2.7 V and $I_{set}$ = 200 pA taken before the line spectroscopy in b). The white line indicates the position of the line spectroscopy cut through the data. b) The average of 3 constant current line spectroscopy cuts through the topography in a) taken $V_{bias}$ = -1.5 and $I_{set}$ = 200 pA is shown.

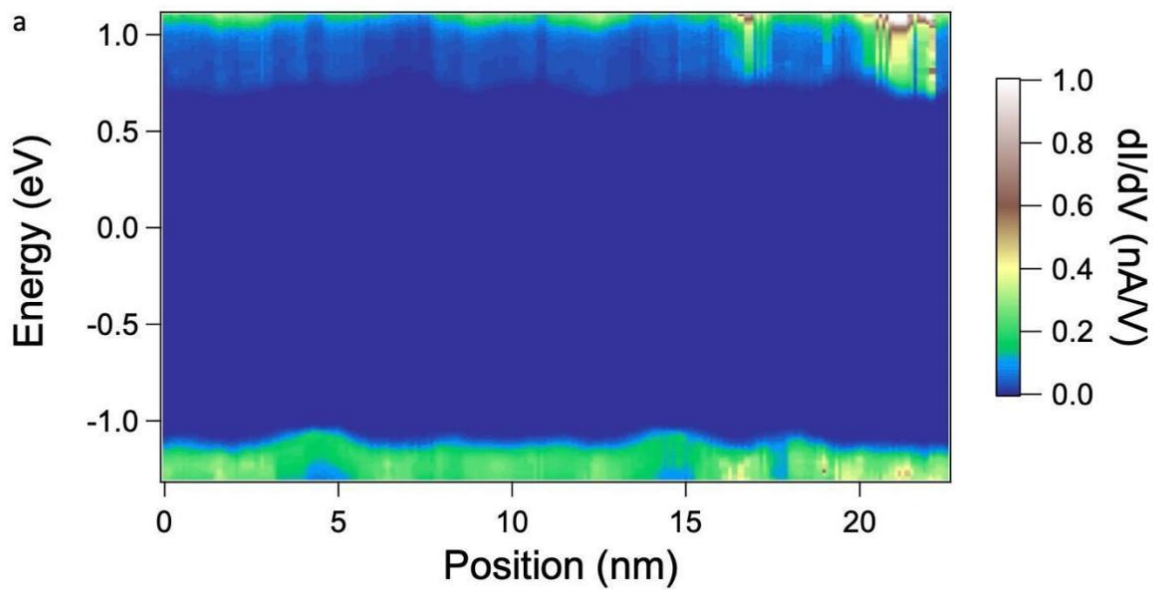

**Supplementary Figure 6.** *Full Band Gap for R-Type Device.* a) Constant Current line spectroscopy taken at $V_{bias}$ = -1.5 V and $I_{set}$ = 200 pA over a range of energy values, -1.5 to 1 eV showing the full band gap of the 5 nm H-type device. The maxima in the valence band do not directly align with the minima or maxima in the conduction band.